\documentclass[reprint,nofootinbib,nobibnotes,amsmath,amssymb,aps,prb,floatfix]{revtex4-1}
\usepackage{epsfig}
\usepackage[dvipsnames,usenames]{color}
\usepackage{color}
\usepackage{amsmath}
\usepackage{amsfonts} 
\usepackage{comment}
\usepackage{array}
\usepackage{multirow}
\usepackage{tabularx}
\usepackage{float}
\usepackage{hyperref}
\usepackage{float}
\usepackage{todonotes}

\newcommand{\pdag}{{\phantom\dagger}}

\newcommand {\apgt} {\ {\raise-.5ex\hbox{$\buildrel>\over\sim$}}\ }
\newcommand {\aplt} {\ {\raise-.5ex\hbox{$\buildrel<\over\sim$}}\ }

\newcommand{\tr}{{\text {Tr} }}
\newcommand{\V}{{\mathcal V}}

\newcommand{\beq}{\begin{equation}}
\newcommand{\eeq}{\end{equation}}
\newcommand{\barray}{\begin{eqnarray}}
\newcommand{\earray}{\end{eqnarray}}
\def\bal#1\eal{\begin{align}#1\end{align}}

\newcommand{\disp}[1]{Eq. (\ref{#1})}

\newcommand{\refdisp}[1]{Ref. [\onlinecite{#1}]}
\newcommand{\figdisp}[1]{Fig. \ref{#1}}

\emergencystretch=\maxdimen
\begin{document}
\title{Pairing correlations in the cuprates: a numerical study of the three-band Hubbard model}
\author{ Peizhi Mai$^{1}$,  Giovanni Balduzzi$^{2}$, Steven Johnston$^{3}$, Thomas A. Maier$^{1}$ }
\affiliation{$^1$Computational Sciences and Engineering Division, Oak Ridge National Laboratory, Oak Ridge, TN, 37831-6494, USA}
\affiliation{$^2$Institute for Theoretical Physics, ETH Zurich, 8093 Zurich, Switzerland }
\affiliation{$^3$Department of Physics and Astronomy, University of Tennessee, Knoxville, Tennessee 37996-1200, USA }
\date{\today}

\begin{abstract}
We study the three-band Hubbard model for the copper oxide plane of the high-temperature superconducting cuprates using determinant quantum Monte Carlo and the dynamical cluster approximation (DCA) and provide a comprehensive view of the pairing correlations in this model using these methods. Specifically, we compute the pair-field susceptibility and study its dependence on temperature, doping, interaction strength, and charge-transfer energy. Using the DCA, we also solve the Bethe-Salpeter equation for the two-particle Green's function in the particle-particle channel to determine the transition temperature to the superconducting phase on smaller clusters. Our calculations reproduce many aspects of the cuprate phase diagram and indicate that there is an ``optimal" value of the charge-transfer energy for the model where $T_c$ is largest. These results have implications for our understanding of superconductivity in both the cuprates and other doped charge-transfer insulators. 
\end{abstract}
\maketitle

\section{Introduction \& Motivation}

Despite decades of active research, the origin of high-temperature (high-$T_c$) superconductivity remains as a central problem in condensed matter physics. Cuprates, perhaps the most studied family of high-$T_c$ superconductors, are experimentally observed to have a $d_{x^2-y^2}$-wave ($d$-wave hereafter) pairing symmetry when their parent compounds are hole- or electron-doped \cite{Kirtley}. The pairing mechanism in cuprates, however, appears to be entirely different \cite{LeeRMP2006, ScalapinoRMP2012} from the one operating in conventional $s$-wave superconductors, which are well described by the BCS theory. Superconductivity in the cuprates occurs in the quasi-two-dimensional CuO$_2$ planes. Here, the half-filled Cu $3d_{x^2-y^2}$ orbitals are highly localized, resulting in a large local on-site Hubbard repulsion and strong electron correlations that are believed to drive emergent superconductivity and a host of unusual normal state behaviors. 
While the prevailing paradigm for understanding these materials is that of a doped Mott insulator~\cite{LeeRMP2006}, the cuprates are properly classified as charge-transfer insulators~\cite{ZSA}. As such, the cuprates have an electron-hole asymmetry in that doped holes (electrons) preferentially reside on the oxygen (copper) sublattice. The Cu and O orbitals of the CuO$_2$ also have a significant degree of hybridization. The minimal model to describe this situation is the three-band Hubbard (or Emery) model \cite{Emery}. 

While the three-band model captures the CuO$_2$ plane's orbital degrees of freedom, it has historically been very challenging to study. The usual way to attack this problem is to map the three-band model onto an effective low-energy single-band (Hubbard or $t-J$) model. This approach assumes that the oxygen orbitals only contribute indirectly to the low-energy sector by establishing the value of the Cu-Cu superexchange $J$ and forming the Zhang-Rice singlet quasiparticles \cite{ZR}. Indeed, the community has obtained a great deal of insight by studying these effective models, and they are believed to contain essential physics of the CuO$_2$ plane. Studies on the single-band Hubbard model, for example, have found that it captures antiferromagnetism in undoped \cite{WhitePRB1989} and lightly doped cases \cite{Dagotto}, $d$-wave superconductivity \cite{Maier2}, high-energy renormalizations in the band dispersion \cite{Macridin2007, Moritz2009}, the presence of a pseudogap \cite{Gull2009}, and the NMR response~\cite{Chen2017}. 

Despite the successes mentioned above, recent studies have also raised new questions. For example, state-of-the-art numerical techniques \cite{Zheng, Qin, JiangYF, JiangHC, Huang, LeBlancPRX} have been used to simulate the single-band Hubbard model with a focus on its ground and low-lying excited states. Collectively, they have found that this sector is characterized by many near-degenerate states, including charge- and spin-orders (i.e. stripes) that compete with superconductivity. One density matrix renormalization group (DMRG) study even concluded that the single-band Hubbard model with only a nearest-neighbor (nn) hopping does not have a superconducting ground state due to competition between these phases \cite{Qin}. The introduction of next-nearest-neighbor (nnn) hopping $t^\prime$, however, can frustrate the stripe order and stabilize $d$-wave superconductivity~\cite{JiangYF}. These results are in conflict with dynamical cluster approximation (DCA) calculations \cite{Maier1, Maier2}, which yield a finite temperature superconducting transition for the Hubbard model with $t^\prime = 0$, but also find that $t^\prime$ is essential to describe the electron-hole asymmetry \cite{Macridin} and experimental band structure \cite{Ogata}. 

These discrepancies underscore two essential points for studying models for the cuprates. First, each numerical method makes its own approximations, which can bias the solution towards one of the many competing low-energy states. It is, therefore, critical to compare results obtained using different numerical methods to get a complete physical picture \cite{LeBlancPRX}. Second, the near degeneracy of the states observed in the single-band model, and their sensitivity to parameters like $t^\prime$, make it necessary to determine whether the downfolding from the three-band to the single-band model introduces other biases. Detailed studies of the three-band Hubbard model are needed to check this.  
Recent advances in high-performance computing make simulations of the three-band Hubbard model more feasible, and several methods have been brought to bear on the problem \cite{Kung, Huang, White, Kent, Hanke,  Biborski,  Weber, Mai}. A detailed determinant quantum Monte-Carlo (DQMC) study \cite{Kung} was conducted on this model to understand the spin-spin and density-density correlations, as well as the spectral function. But given the discrepancies between methods observed for the single-band model, it is also essential to attack this problem using DCA~\cite{Maier1}. DCA can access the thermodynamic limit in 2D, and is, therefore, able to resolve the superconducting instability signalled by a divergence of the pair-field susceptibility~\cite{Maier2}. This capability is not present in finite-size methods 
like DQMC or DMRG. 

\begin{figure}[b]
\center{\includegraphics[width=.75\columnwidth]{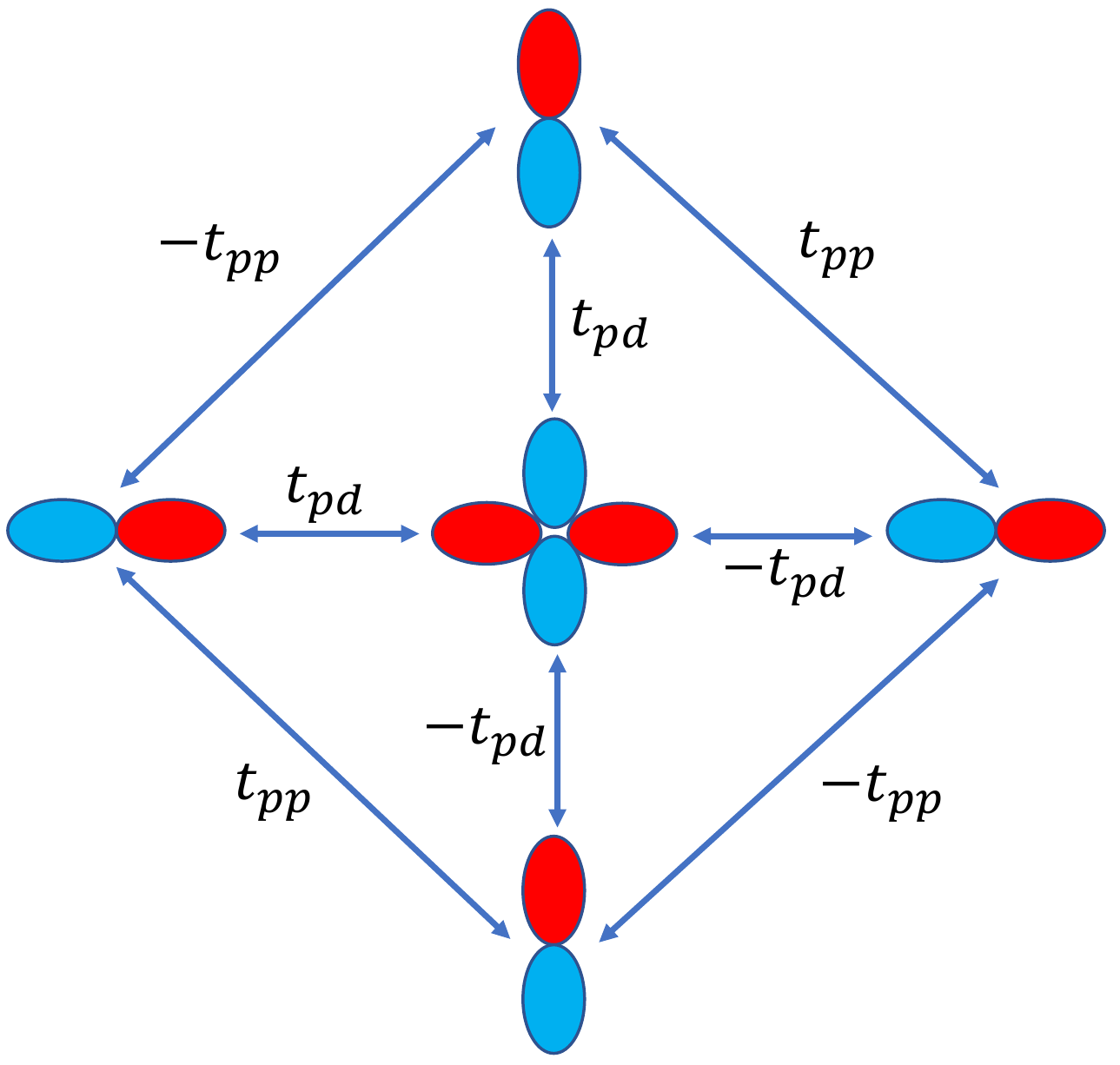}}
\caption{ The orbital basis for the three-band Hubbard model includes a cooper $d_{x^2-y^2}$ orbital and its surrounding oxygen $p_x$ and $p_y$ orbitals. The convention of the hopping integrals is shown in the figure.
}
\label{threeband}
\end{figure}

With these motivations, we previously studied the three-band Hubbard model within the DCA~\cite{Mai}. By solving the Bethe-Salpeter equation, we showed that the pairing interaction in the three-band model has a simple $d$-wave pairing pattern with dominant weights located on the Cu $d$ and oxygen bonding molecular orbitals. This finding enabled us to unambiguously define and calculate the correct orbital-dependent $d$-wave pairing susceptibility. This aspect has been lacking in earlier studies, which have each adopted different definitions for the pairing susceptibility and failed to reach a consensus regarding pairing correlations~\cite{Scalettar, Guerrero,  Moreo, Biborski}. Based on this recent progress, here we use DQMC and DCA to study the $d$-wave pairing correlations of the three-band Hubbard model more broadly. We first establish an agreement between the susceptibility measured by both methods at high temperature and the marginal effect of on-site interaction in the oxygen orbitals. Then we present DCA results for the superconducting $T_c$ obtained on small clusters and argue that $T_c$ is better predicted by the solutions to the Bethe-Salpeter equation (BSE) than by the pair-field susceptibility at temperatures well above $T_c$. By varying the charge-transfer energy, we also demonstrate that an ``optimal" $T_c$ occurs for a fixed hole-density and discuss the reasons for this behavior. 

\section{Model and Parameters}

The three-band Hubbard model's Hamiltonian is defined 
as $H=K+V_{dd}+V_{pp}$, where $K=K_0+K_{pd}+K_{pp}$ and 
\begin{eqnarray}\nonumber
K_0&=&(\varepsilon_d-\mu)\sum_{i,\sigma}n_{i,\sigma}^d+ (\varepsilon_p-\mu)\sum_{i,\alpha,\sigma} n^p_{i,\alpha,\sigma} \\\nonumber
K_{pd}&=&\sum_{\langle  i,j,\alpha\rangle,\sigma}t^{i,j,\alpha}_{pd}(d^\dagger_{i,\sigma}p^\pdag_{j,\alpha,\sigma}+p^\dagger_{j,\alpha,\sigma}d^\pdag_{i,\sigma})
\\\nonumber
K_{pp}&=&\sum_{\langle j,\alpha,j^\prime,\alpha^\prime \rangle,\sigma} t^{j,j^\prime,\alpha,\alpha^\prime}_{pp}(p^\dagger_{j,\alpha,\sigma}p^\pdag_{j^\prime,\alpha^\prime,\sigma}+p^\dagger_{j^\prime,\alpha^\prime,\sigma}p^\pdag_{j,\alpha,\sigma})
\\\nonumber V_{dd}&=&U_{dd} \sum_i n^d_{i,\uparrow}n^d_{i,\downarrow}
\\V_{pp}&=&U_{pp} \sum_{j,\alpha} n^p_{j,\alpha,\uparrow}n^p_{j,\alpha,\downarrow}.
\end{eqnarray}
Here, $d^\dagger_{i,\sigma}$ ($d^\pdag_{i,\sigma}$) creates (annihilates) a spin $\sigma$ ($=\uparrow,\downarrow$) hole in the copper $d_{x^2-y^2}$ orbital at site $i$;  $p^\dagger_{i,\alpha,\sigma}$ ($p^\pdag_{i,\alpha,\sigma}$) creates (annihilates) a spin  $\sigma$ hole in the oxygen $p_\alpha$ ($\alpha = x,y$) orbital at site $j$; $\langle \dots \rangle$ denotes a sum over nearest-neighbor orbitals; $n^d_{i,\sigma}=d^\dagger_{i,\sigma}d^\pdag_{i,\sigma}$ and 
$n^p_{i,\alpha,\sigma}=p^\dagger_{i,\alpha,\sigma}p^\pdag_{i,\alpha,\sigma}$ are the number operators; 
$\epsilon_d$ and $\epsilon_p$ are the on-site energies of the Cu and O orbitals, respectively; $\mu$ is the chemical potential; $t^{i,j,\alpha}_{pd}$ and $t^{j,j^\prime,\alpha,\alpha^\prime}_{pp}$ are the nearest neighbor Cu-O and O-O hopping integrals; and $U_{dd}$ and $U_{pp}$ are the on-site Hubbard repulsion on the Cu and O orbitals, respectively. 

The hopping integrals are parameterized as $t^{i,j,\alpha}_{pd} = P_{i,j,\alpha}t_{pd}$ and $t^{j,j^\prime,\alpha,\alpha^\prime}_{pp} =  Q_{j,j^\prime,\alpha,\alpha^\prime} t_{pp}$, where $P_{i,j,\alpha}$ and $Q_{j,j^\prime,\alpha,\alpha^\prime}$ take values $\pm 1$ following the convention shown in \figdisp{threeband}. Throughout this work, we adopted a canonical parameter set for the cuprates (in units of eV): $t_{pd} = 1.13$, $t_{pp} = 0.49$, $U_{dd} = 8.5$, $U_{pp} = 0$, and $\Delta = \varepsilon_p -\varepsilon_d = 3.24$~\cite{Kung, Czyzyk, Johnston, Ohta}, unless otherwise stated. Since we use a hole language,  half-filling is defined as a hole density $n_h =1$, which means one hole per unit cell. $n_h>1$ then corresponds to hole-doping and $n_h<1$ corresponds to electron-doping.

In our recent DCA study \cite{Mai}, we observed a pairing interaction with a simple $d$-wave structure that exists between the Cu $d$ and oxygen bonding and anti-bonding molecular orbitals, denoted here 
as $p_L$ and $p_{L^\prime}$, respectively. The unitary transformation \cite{ZR,Avella2013,Maier4} from the oxygen $p_x$, $p_y$ orbital basis to $p_L$, $p_{L^\prime}$ molecular orbital basis is defined in $k$-space as 
\begin{equation}
p_{L,{\bf k},\sigma}=\frac{\mathrm{i}}{\gamma_{{\bf k}}} \left[\sin\left(\tfrac{k_xa}{2}\right) p_{x,{\bf k},\sigma}- \sin\left(\tfrac{k_ya}{2}\right) p_{y,{\bf k},\sigma}\right]\label{Lk}
\end{equation}
and
\begin{equation}
p_{L^\prime,{\bf k},\sigma}=-\frac{\mathrm{i}}{\gamma_{{\bf k}}} \left[\sin\left(\tfrac{k_ya}{2}\right) p_{x,{\bf k},\sigma}+ \sin\left(\tfrac{k_xa}{2}\right) p_{y,{\bf k},\sigma}\right],\label{Lbark}
\end{equation}
where $\gamma^2_{{\bf k}}=\sin^2(k_xa/2)+\sin^2(k_ya/2)$,  $p_{\alpha,{\bf k},\sigma}=N^{-1/2}_c\sum_{j} p_{\alpha, j ,\sigma}\exp(-\mathrm{i} {\bf k} \cdot {\bf R}_j)$, and the lattice constant $a$ is set to $1$. A feature of this basis is that the $d$ and $p_{L^\prime}$ orbitals only hybridize with the $p_L$ state and not with each other. The Fourier transform of the $p_L$ and $p_{L^\prime}$ orbitals to real-space is defined as $p_{L,i,\sigma}=N^{-1/2} \sum_{\bf k}p_{L,{\bf k},\sigma} \exp(-\mathrm{i} {\bf k} \cdot {\bf R}_i)$, $p_{L^\prime,i^\prime,\sigma}=N^{-1/2} \sum_{\bf k}p_{L^\prime, {\bf k},\sigma} \exp(-\mathrm{i} {\bf k} \cdot {\bf R}_{i^\prime})$, where $i^\prime=i+\hat{x}/2+\hat{y}/2$.

\vskip 0.5cm 
\noindent
\section{Methods}
\subsection{Determinant Quantum Monte Carlo}  
We first provide a brief overview of the determinant quantum Monte Carlo (DQMC) algorithm. Additional details also can be found in Refs. [\onlinecite{Dopf1, Dopf2, Scalettar, Kung}].  

DQMC is a non-perturbative auxiliary field technique that computes the expectation values of an observable in the grand canonical ensemble
\beq
\langle\hat{O}\rangle =
\frac{1}{\mathcal{Z}}\tr\left[\hat{O} \ \text{e}^{-\beta H}\right],\label{Eq:Oexp}
\eeq
where $\mathcal{Z} = \tr\left[ \text{e}^{-\beta H}\right]$ is the partition function. To evaluate Eq.~(\ref{Eq:Oexp}), the imaginary-time interval $\left[0,\beta\right]$ is divided into $L$ evenly spaced slices of width $\Delta\tau=\tfrac{\beta}{L}$. Once this is done, the exponential is decomposed using the Trotter approximation such that
\beq\nonumber
\text{e}^{-L\Delta\tau H} \approx ( \text{e}^{-\Delta\tau K} \text{e}^{-\Delta\tau V_{dd}} \text{e}^{-\Delta\tau V_{pp}})^L.
\eeq

A trace over the quadratic terms can be evaluated directly \cite{BSS}. The quartic interacting terms are, therefore, transformed into a quadratic form by introducing a Hubbard-Stratonovich transformation
\beq\nonumber
\begin{split}
&\text{e}^{-\Delta\tau U_{\alpha\alpha}n^{\alpha}_{i,\uparrow}n^{\alpha}_{i,\downarrow}} \\&= \frac{1}{2}
\sum_{s_{i,\alpha,l}} s_{i,\alpha,l} 
e^{{\lambda_\alpha s_{i,\alpha,l} (n_{i\uparrow}^{\alpha}-n_{i\downarrow}^{\alpha})-\frac{1}{2}\Delta\tau U_{\alpha\alpha} (n^{\alpha}_{i\uparrow}+n^{\alpha}_{i\downarrow})}},
\end{split}
\eeq
where we introduce auxiliary Hubbard-Stratonovich fields $s_{i,\alpha,l}=\pm 1$ at each space-(imaginary)time point $(i,l)$, $\alpha$ is an orbital index, and $\lambda_{\alpha}$ is defined by $\tanh^2(\lambda_\alpha/2)=\tanh(\Delta\tau U_{\alpha\alpha}/4)$.  

Once the interacting terms are rewritten in a quadratic form, we can evaluate the trace over the Fermionic degree of freedom to obtain an expression for the partition function in terms of matrix determinants 
\begin{equation*}
    Z = \sum_{s_{i,\alpha,l}=\pm 1} \det M^{+} \det M^{-}, 
\end{equation*}
where
$M^{\alpha}=I + B^{\sigma}_LB^{\sigma}_{L-1}...B^{\sigma}_{1}$,
$B^{\pm}_l= e^{-\Delta\tau K} e^{v^d_{\pm}(l)}e^{v^p_{\pm}(l)}$, 
$I$ is the identity matrix, 
and $v^\alpha_{\pm}(l)$ are matrices whose elements are given by
\beq\nonumber
v^{\alpha}_{\pm}(l)_{mm^\prime}=\delta_{mm^\prime}\left[\pm \lambda_{\alpha}s_{m,\alpha,l} - \Delta\tau\frac{U_{\alpha\alpha}}{2} \right].
\eeq
We can then calculate the expectation values of an observable $\langle\hat{O}\rangle$ by sampling the Hubbard-Stratonovich fields using the Markov-chain Monte Carlo method, where the system accepts proposed local and global changes using a modified heat bath algorithm.

The weight of each Hubbard-Stratonovich field configuration is given by $W(\{s_{i,\alpha,l}\}) = \tfrac{1}{Z}\det M^{+} \det M^{-}$, which is not positive definite. This aspect is a manifestation of the infamous fermion sign problem. To deal with it, we separate the weight into a ``probability" $P(s)$ representing its absolute value of $\det M^{+} \det M^{-}$ and $f_\mathrm{sign}$ representing its sign. ($f_\mathrm{sign}$ is  commonly referred to as the Fermion sign.) The expectation values are then re-weighted as:
\beq\nonumber
\langle\hat{O}\rangle =\frac{\sum_{s_{m,\alpha,l}} \hat{O} f_\mathrm{sign} P(s)}{\sum_{s_{m,\alpha,l}} f_\mathrm{sign} P(s)},
\eeq
where the denominator measures the average value of the fermion sign $\langle f_\mathrm{sign}\rangle$. The average sign is usually less than one except for some special cases where it is protected by its symmetry (e.g. the half-filled single-band Hubbard model) \cite{Iglovikov2015}. In general, $\langle f_\mathrm{sign}\rangle$ decreases with increasing lattice size or decreasing temperature. When $\langle f_\mathrm{sign}\rangle$ is close to zero, the statistical fluctuations in measuring an observable will be magnified and many more measurements are needed to obtain an accurate result.

\vskip 0.5cm 
\noindent
\subsection{ The Dynamical Cluster Approximation}\label{sec:MethodsDCA}
The dynamical cluster approximation (DCA)~\cite{Maier1, Maier2, Maier3} maps the bulk lattice in the thermodynamic limit to a finite size cluster embedded in a self-consistent mean-field that approximates the remainder of the system. DCA describes a system in the thermodynamic limit by treating the short-range correlations within the cluster explicitly, and the longer-range correlations as a dynamical mean-field. An interested reader can find further details about the DCA method in Ref.~\cite{Maier1}. Here, we provide an overview and highlight the aspects that are needed to treat the three-band Hubbard model.

The basic assumption of the DCA is that the dominant correlations are  primarily short-ranged and can be captured within the cluster. With that, the self-energy $\Sigma_{\alpha_1,\alpha_2}({\bf k},\text{i} \omega_n)$ is approximated by $\Sigma_{\alpha_1,\alpha_2}({\bf K},\text{i} \omega_n)$, where ${\bf K}$ are the cluster momenta, and $\alpha_1$ and $\alpha_2$ are band indices. The coarse-grained single-particle Green's function can be obtained by 
\begin{widetext}
\beq
\bar{G}_{\alpha_1,\alpha_2}({\bf K},\text{i}\omega_n)=\frac{N_c}{N}\sum_{\bf{k}}G_c({\bf K}+{\bf k},\text{i}\omega_n)_{\alpha_1,\alpha_2}=\frac{N_c}{N}\sum_{\bf{k}}\left[(\text{i}\omega_n+\mu)I-\varepsilon({\bf K}+{\bf k})-\Sigma({\bf K},\text{i}\omega_n)\right]^{-1}_{\alpha_1,\alpha_2},
\eeq
where $\mu$ is the chemical potential, which adjusted to obtain a given density, and $N_c$ is the number of unit cells in the cluster. Here, the dispersion $\varepsilon({\bf K+k})$ is a 3$\times$3 matrix in the $d$, $p_x$ and $p_y$ orbital basis, obtained by Fourier-transforming the hopping integrals in Eq.~(1). The coarse-grained sum is over the momenta $\bf{k}$ in a square patch centered at ${\bf K}$ whose size is determined by the ratio of the Brillouin zone volume to the size of the cluster. This procedure reduces the bulk problem to a finite size cluster problem, which we solve self-consistently using the continuous-time auxiliary field quantum Monte-Carlo algorithm \cite{Gull}. 

To study the pairing correlations in the normal state, we solve the Bethe-Salpeter equation (BSE)
\begin{equation} \label{eq:BSE}
    -\frac{T}{N_c} \sum_{K^\prime,\alpha_1,\alpha_2} \Gamma^{c, pp}_{\alpha,\beta,\alpha_1,\alpha_2}(K,K^\prime)\bar{\chi}_{\alpha_1,\alpha_2,\alpha_3,\alpha_4}(K^\prime) \phi^{R,\nu}_{\alpha_3\alpha_4}(K^\prime) = \lambda_\nu \phi^{R,\nu}_{\alpha\beta}(K)\,.
\end{equation}
Here, $K=({\bf K},\text{i}\omega_n)$, and $\bar{\chi}_{\alpha_1,\alpha_2,\alpha_3,\alpha_4}({\bf K}, \text{i}\omega_n) = (N_c/N) \sum_{{\bf k}^\prime} [G_{\alpha_1\alpha_3}({\bf K+k'},\text{i}\omega_n)G_{\alpha_2\alpha_4}(-{\bf K}-{\bf k}^\prime, -\text{i}\omega_n)]$ is the coarse-gained bare particle-particle propagator. The irreducible particle-particle vertex $\Gamma^{c, pp}_{\alpha_1,\dots,\alpha4}(K,K')$ is assumed to only depend on the cluster momenta ${\bf K}$. It is extracted from the two-particle cluster Green's function $G^{2,c}_{\alpha_1,\dots,\alpha_4}(K,K')$ with zero center of mass momentum and frequency by inverting the cluster Bethe-Salpeter equation
\begin{eqnarray}\nonumber
G^{2,c}_{\alpha_1,\alpha_2,\alpha_3,\alpha_4}(K,K^\prime) &=&\bar{G}_{\alpha_1,\alpha_3}(K)\bar{G}_{\alpha_2,\alpha_4}(-K)\delta_{K,K^\prime} \\
&+&\frac{T}{N_c}\sum_{\substack{K^{\prime\prime}\\\alpha^\prime_1,\dots,\alpha^\prime_4}}\bar{G}_{\alpha^{\phantom\prime}_1,\alpha^\prime_1}(K)\bar{G}_{\alpha^{\phantom\prime}_2,\alpha^\prime_2}(-K)
\Gamma^{c,pp}_{\alpha^\prime_1,\alpha^\prime_2,\alpha^\prime_3,\alpha^\prime_4}(K,K^{\prime\prime})G^{2,c}_{\alpha^\prime_3,\alpha^\prime_4,\alpha^{\phantom\prime}_3,\alpha^{\phantom\prime}_4}(K^{\prime\prime},K^\prime)\,.
\label{BSE1}
\end{eqnarray}
\end{widetext}

To keep left and right eigenvectors of the eigenvalue equation (\ref{eq:BSE}) consistent, we symmetrize the pairing kernel entering Eq. (\ref{eq:BSE}). Using matrix notation in $(K, \alpha, \beta)$, we first diagonalize the bare particle-particle propagator, $\bar{\chi}^D = U^{-1}\bar{\chi}U$, where $\chi^D$ is a diagonal matrix. We then use the 
transformation matrix $U$ to symmetrize the BSE
\begin{equation}\label{sBSE}
-\frac{T}{N_c} U \sqrt{\chi^D}U^{-1} \Gamma^{c,pp} U \sqrt{\chi^D}U^{-1} \phi^\nu = \lambda_\nu \phi^\nu\,.
\end{equation}
We use the eigenvectors of the symmetrized BSE  $\phi^\nu_{\alpha\beta}(K)$ for our analysis presented here. 
They are related to the right eigenvectors of the BSE in 
Eq.~(\ref{eq:BSE}) by 
\begin{equation}\nonumber
    \phi^\nu = U\sqrt{\chi^D}U^{-1}\phi^{R,\nu}\,.
\end{equation}

The coarse-grained two-particle Green's 
function is defined as 
\beq
\bar{G}_2(K,-K,-K^\prime,K^\prime) = \frac{N_c^2}{N^2}\sum_{k,k^\prime}G^{2,c}(k,-k,-k^\prime,k^\prime).
\label{cgG2}
\eeq
It can be constructed from the eigenvalues and eigenvectors using \disp{BSE1}
\beq
\begin{split}
&\bar{G}_{2}=
[U\sqrt{\chi^D}U^{-1}]\sum_{\nu}\frac{[\phi^{-1}]^{\nu}\phi^{\nu}}{1-\lambda_{\nu}}[U\sqrt{\chi^D}U^{-1}].\label{G2r}
\end{split}
\eeq
This equation shows that $\bar{G}_2$ diverges as the leading eigenvalue $\lambda_{\nu}\rightarrow 1$. The temperature at which this divergence occurs 
is the superconducting transition temperature $T_c$.  

Our DQMC and DCA calculations are both limited by the fermion sign problem. The coupling of the cluster to a self-consistent mean-field in DCA reduces the sign problem considerably, however, which allows us to access physics at lower temperatures~\cite{Maier1, Jarrell}.

\section{Results \& Discussion}
In this section, we define the components of the $d$-wave pairing susceptibility in the  $\{d,p_L,p_{L^\prime}\}$ basis following our recent DCA study \cite{Mai}. We then present the filling and charge-transfer (CT) energy dependence of the pair-field susceptibility and transition temperature from both DCA and DQMC. 

\subsection{The $d$-wave pair-field susceptibility}
We examine several components of the $d$-wave pair-field susceptibility defined in the $\{d,p_L,p_{L^\prime}\}$ basis. We previously have shown that in this basis, all the contributions have $d$-wave symmetry \cite{Mai}. 
First, we define a generalized pair-field susceptibility in the $d$-wave channel 
\beq
P_{d,\alpha_1\alpha_2\alpha_3\alpha_4}=\int_0^{\beta} d\tau\langle \Delta^{\dagger}_{\alpha_1,\alpha_2}(\tau) \Delta^\pdag_{\alpha_3,\alpha_4}(0) \rangle, \label{Pdall}
\eeq
where $\alpha_i=d$, $p_L$, $p_{L^\prime}$ are orbital indices,  $\Delta^{\dagger}_{\alpha_1,\alpha_2}=\frac{1}{\sqrt{N}}\sum_{\bf k}g^{\phantom\dagger}_d(\bf{k})c^{\dagger}_{\alpha_1,\bf{k},\uparrow}c^{\dagger}_{\alpha_2,-\bf{k},\downarrow}$, $g_d({\bf k}) = \cos(k_x)-\cos(k_y)$, and $c^{\dagger}_{\alpha,\bf{k},\sigma}$ creates a hole in orbital $\alpha$ with momentum ${\bf k}$ and spin $\sigma$.
The total $d_{x^2-y^2}$-wave pair-field susceptibility is defined as summation over all channels:

\begin{figure}[t!]
\centering
\includegraphics[width=0.8\linewidth]{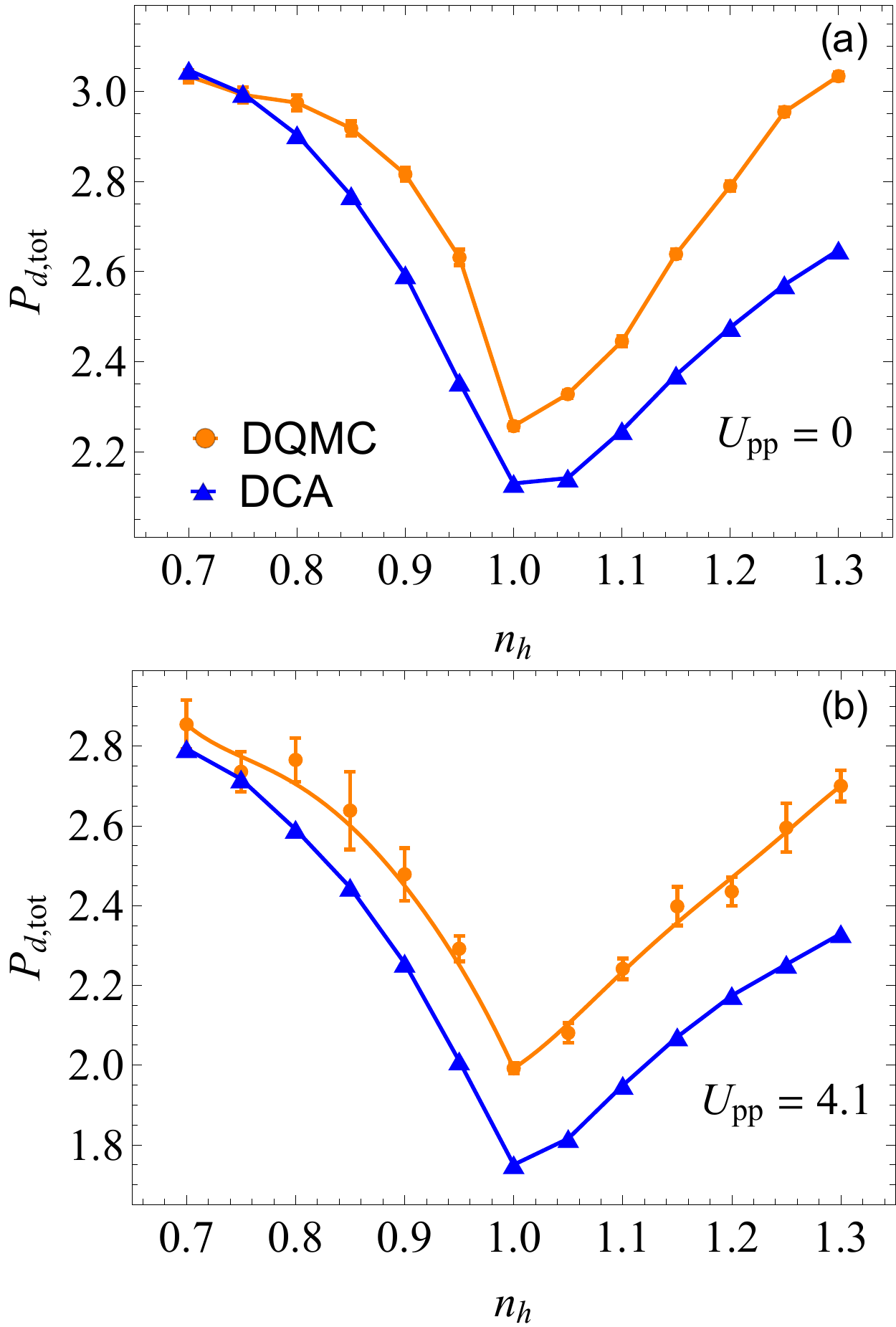}
\caption{The total $d$-wave pair-field susceptibility $P_{d,\text{tot}}$ as a function of hole density from DQMC and DCA at $t_{pd}=1.13$eV, $t_{pp}=0.49, \Delta=3.24, U_{dd}=8.5, T=0.125, N_{\textrm{Cu}}=4\times 4, \Delta\tau=0.1$ (DQMC) with $U_{pp}=0$ and $U_{pp}=4.1$. The two methods largely agree. In both cases, 
setting $U_{pp}\ne 0$ changes $P_{d,\text{tot}}$ only slightly while also increasing the statistical error bars. The latter effect 
is due to a decrease in the average value of the fermion sign. The lines in all panels are guides for the eye.
}
\label{Upp}
\end{figure}

\beq
P_{d,\text{tot}}=\sum_{\substack{\alpha_1,\alpha_2,\\\alpha_3,\alpha_4}} P_{d,\alpha_1\alpha_2\alpha_3\alpha_4}.
\eeq
This quantity characterizes how easily Cooper pairs form once an infinitesimal pair-field switches on, and provides a measure of the superconducting correlations in the normal state.

\begin{figure}[ht]
\centering
\includegraphics[width=0.83\linewidth]{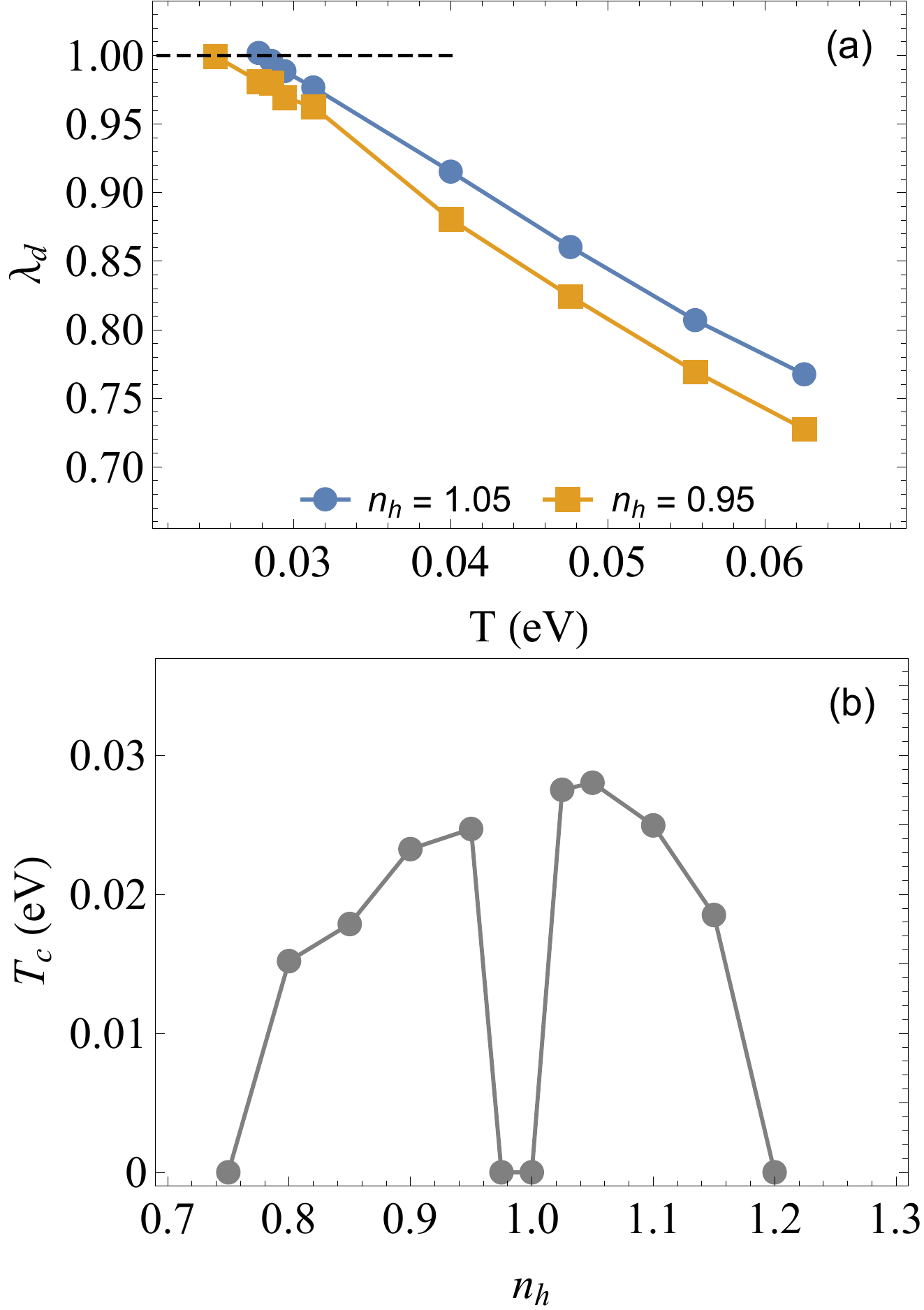}
\caption{ (a) $\lambda_d$ vs $T$ at different hole densities. Using DCA, we were able to reach low enough temperature where $\lambda_d$ goes beyond $1$ and estimate $T_c$ by interpolation. (b) $T_c$ versus $n_h$ shows two superconducting domes on the hole-doped and electron-doped sides, respectively, with the maximum $T_c$ higher on the hole-doped side, consistent qualitatively with experiments. The peaks are located around $0.05$ doping close to half-filling due to the small cluster size. 
}
\label{Tcdopedepend}
\end{figure}

\begin{figure*}[ht]
\centering
\includegraphics[width=0.8\linewidth]{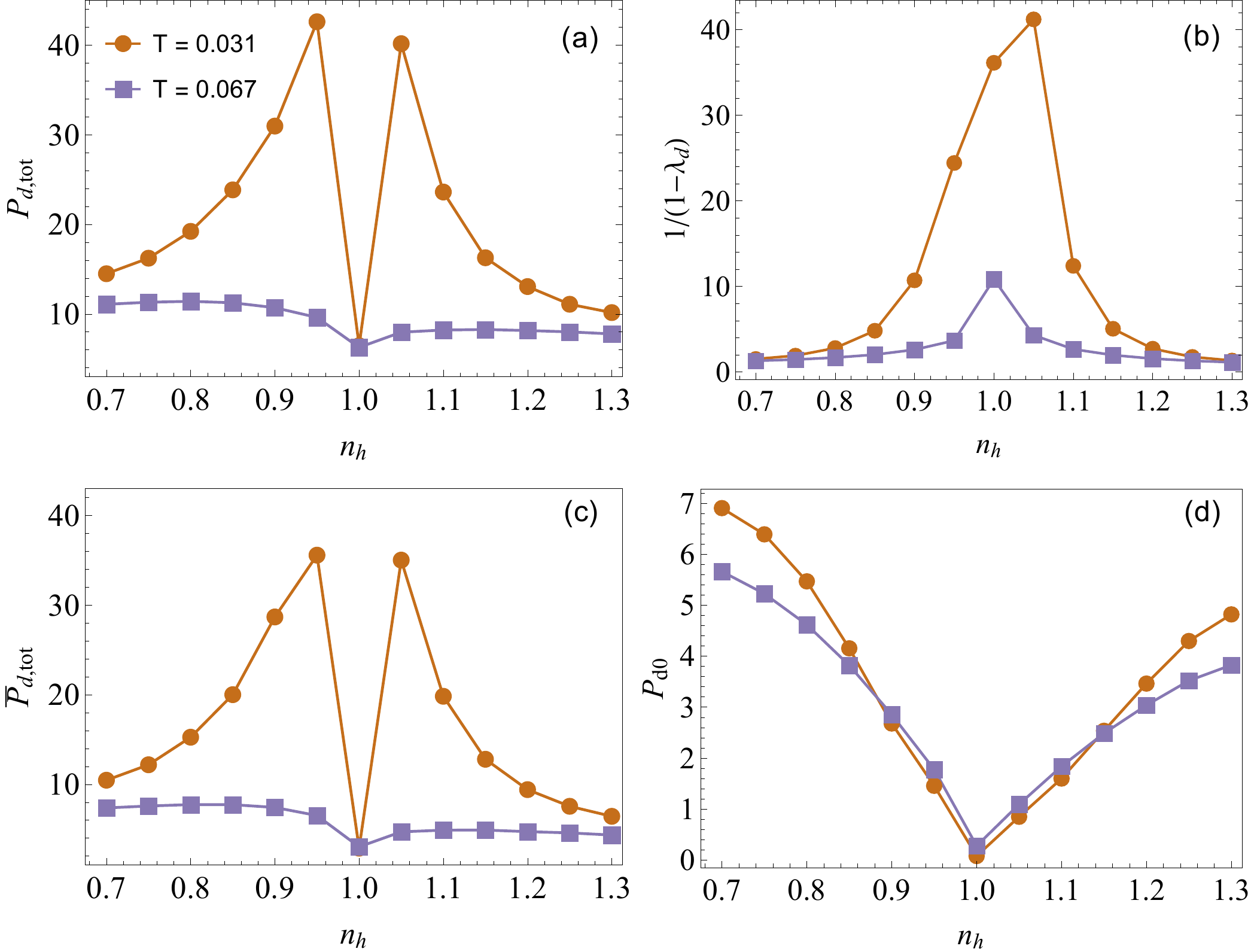}
\caption{(a) The pair-field susceptibility and its factors (b) $1/(1-\lambda_d)$, (c) the approximate pair-field susceptibility 
$\bar{P}_{d,\text{tot}}$, and (d) the pre-factor $P_{d0}$, all plotted 
as functions of hole density. Results shown here were obtained from a DCA calculation on a $2\times2$ cluster. Each panel shares the same legend. The asymmetry of the pair-field susceptibility is determined by the intrinsic pair-field susceptibility $P_{d0}$, while the divergence of $1/(1-\lambda_d)$ determines $T_c$. The lines in all panels are guides for the eye.
}
\label{Pddopedepend}
\end{figure*}

\begin{figure}[ht]
\includegraphics[width=0.8\linewidth]{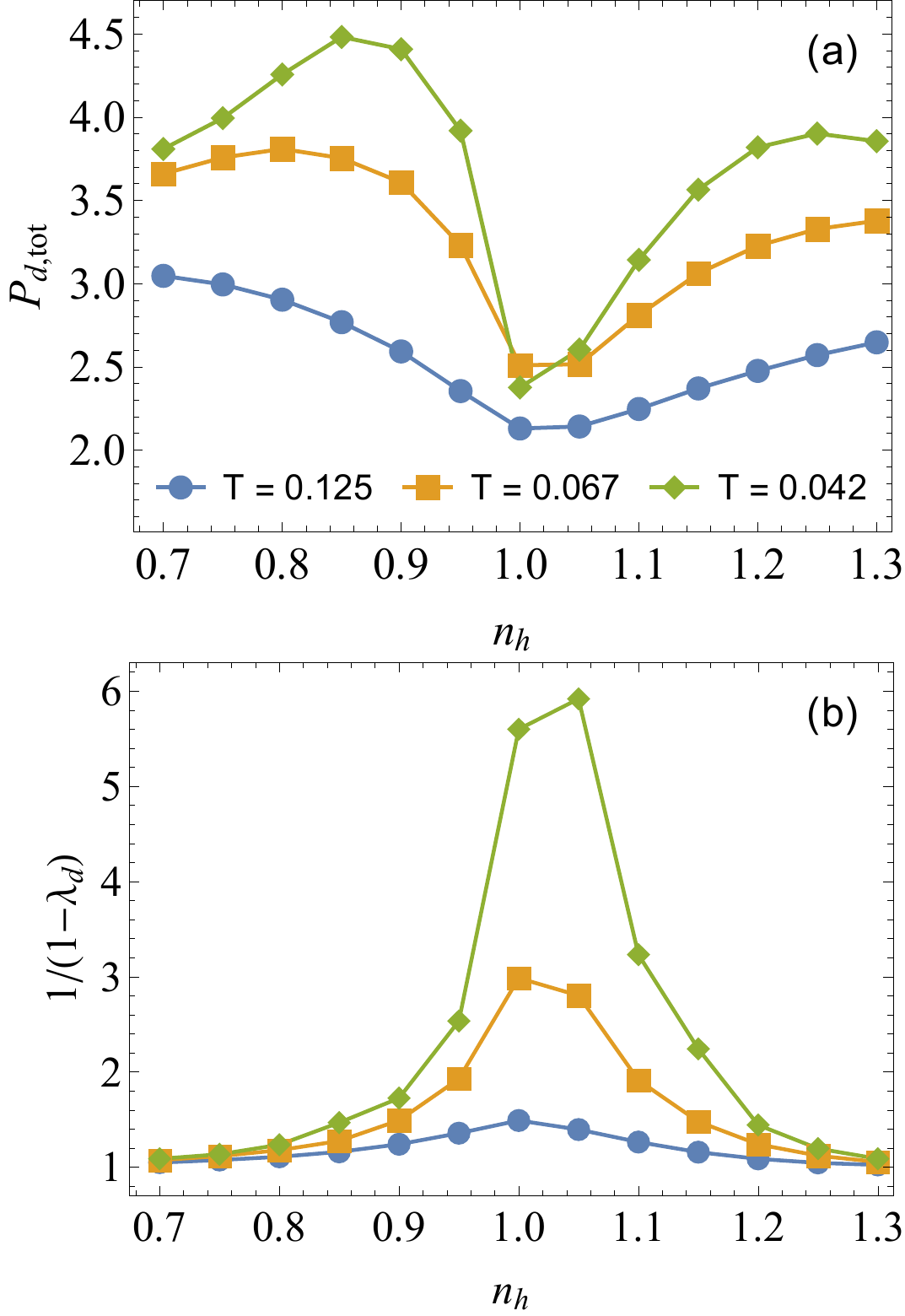}
\caption{(a) The pair-field susceptibility and its factor $1/(1-\lambda_d)$  (b) vs. hole density, obtained from a DCA calculation on a $4\times 4$ cluster for different temperatures. Each panel shares the same legend. At the lowest temperature, the magnitude of $1/(1-\lambda_d)$ is larger on the hole-doped side, just like $T_c$ in the 2$\times$2 cluster, while the pair-field susceptibility shows the opposite asymmetry. The lines are guides for the eye.
}
\label{dopedepend4by4}
\end{figure}

\subsection{High-temperature pairing correlations} 
Figure \ref{Upp} presents DQMC and DCA results for $P_{d,\text{tot}}$ 
as a function of hole density at $T=0.125$ eV$^{-1}$ obtained on an $N_{\textrm{Cu}}=4\times4$ cluster with $U_{pp}=0$ eV [\figdisp{Upp}(a)] and $U_{pp}=4.1$ eV [\figdisp{Upp}(b)]. 
DQMC treats the problem with a finite-size cluster in a numerically exact way, while DCA embeds its cluster self-consistently in a dynamical mean-field. We, therefore, expect the two methods to give slightly different results that converge towards one another as the cluster size increases. 
Fig. \ref{Upp} shows that, at high temperatures and for the 16-site cluster, both DQMC and DCA already give very similar results. Moreover, comparing the results in \figdisp{Upp}(a) and \figdisp{Upp}(b), we find that setting $U_{pp}=4.1$ eV marginally effects $P_{d,\text{tot}}$ while also increasing the statistical error bars. We can rationalize this observation by recalling that $U_{pp}$ increases the correlations in the doped system and exacerbates the sign problem~\cite{Kung}. For this reason, we will set it to zero for the remainder of this study. 

The agreement between DQMC and DCA, two independent numerical methods, gives us confidence that the three-band model has been implemented correctly in the DCA++ code, and shows that both methods produce very similar results when correlations are short-ranged and sufficiently represented within the cluster. We note, however, that DCA approximates the thermodynamic limit by treating long-range correlations at the mean-field level. At the same time, DQMC solves the problem exactly on a finite-size cluster (with the associated finite-size effects). For these reasons, we expect that we would see more significant differences between the two methods at lower temperatures (if the sign problem were not present) when considering small clusters like those considered here. These differences, however, would vanish as the cluster size increases.

Inspecting \figdisp{Upp}, we find that the total pair-field susceptibility is larger on the electron-doped side compared to the hole-doped side for the fillings we investigated, which is contrary to expectations. Upon examining the dominant components of this quantity (shown in the Appendix), we observe that all of them exhibit a similar asymmetry between electron- and hole-doping. This difference arises because the doped holes tend to go to the O orbitals while doped electrons tend to go to the Cu orbitals. We don't believe that the larger susceptibility on the electron-doped side is a finite-size effect, since we observe similar asymmetries in a $2\times2$ cluster, as discussed in the next section. One might naively expect the reverse asymmetry since the experimental transition temperature is higher for the hole-doped cuprates. However, as we will show in the next section, the high-temperature pairing susceptibility can be a poor proxy for the actual $T_c$ realized in the system.

\subsection{Density dependence of pair-field susceptibility and transition temperature}\label{Tc}
Our DCA calculations are able to reach temperatures low enough to directly determine the superconducting $T_c$ on a $2\times2$ cluster, the minimum size needed to support a $d$-wave symmetry. To demonstrate this, 
\figdisp{Tcdopedepend}(a) plots $\lambda_d$ vs $T$ for a hole- ($n_h=1.05$) and electron-doped ($n_h=0.95$) system. As discussed in Sec.~\ref{sec:MethodsDCA}, the superconducting transition occurs when $\lambda_d(T_c)=1$. Here, we are able to track $\lambda_d$ \textit{across} this value so that we can extract $T_c$ by  interpolating the data. 
Using this procedure, we are able to explicitly compute $T_c$ as a function of doping, as shown in \figdisp{Tcdopedepend}(b). The results shown in \figdisp{Tcdopedepend}(b) display a superconducting dome on both the hole- and electron-doped sides with the maximum $T_c=0.028$~eV ($\beta\approx 36$) at $n_h=1.05$. These observations are consistent with the those made in a previous DCA study on a two-band Hubbard model, where $T_c$ was obtained from an extrapolation of the $d$-wave pair-field susceptibility  \cite{Macridin}. 

Our $T_c$ results on the $2\times2$ cluster capture many qualitative aspects of the cuprate phase diagram. For example, $T_c^\mathrm{max}$ is larger on the hole-doped side, and the superconducting dome is wider on the electron-doped side in comparison to the hole-doped side; however, we also obtain $T_c$ values that are larger than those observed experimentally~\cite{Tallon}, and our ``optimal" doping values appear closer to half-filling. We believe that both inconsistencies are due to the use of a small 2$\times$2 cluster or possibly additional physics not included in the model (e.g.  inhomogeneities). For the single-band Hubbard model, previous DCA calculations have found that $T_c$ decreases considerably when larger clusters are considered \cite{Maier2}. 

Next, we determine how the particle-hole asymmetry in $T_c$ is related to that of $P_{d,\text{tot}}$ found above. This question is imperative since the pair-field susceptibility is often used as a proxy for the superconducting transition \cite{Maier2, Scalettar, Biborski,  Macridin}. \figdisp{Pddopedepend} presents an analysis of the pair-field susceptibility, this time obtained on a $2\times 2$ cluster to facilitate a direct comparison to \figdisp{Tcdopedepend}. 
As with the $4\times4$ case, $P_{d,\text{tot}}$ has a reversed asymmetry compared to $T_c$, in that it is significantly higher on the electron-doped side. 
To understand this discrepancy better, we now analyze $P_{d,\text{tot}}$ in more detail.

According to Eqs.~(\ref{G2r}) and (\ref{Pdall}), we can write $P_{d,\text{tot}}$ as
\beq
\begin{split}
P_{d,\text{tot}}=g_d
[U\sqrt{\chi^D}U^{-1}]\sum_{\nu}\frac{[\phi^{-1}]^{\nu}\phi^{\nu}}{1-\lambda_{\nu}}[U\sqrt{\chi^D}U^{-1}]g_d.\label{Pdall2}
\end{split}
\eeq
The contribution from the leading $d$-wave eigenvector and eigenvalue becomes dominant at low temperature, leading to an approximate pair-field susceptibility:
\beq
\begin{split}
\bar{P}_{d,\text{tot}}&=g_d
[U\sqrt{\chi^D}U^{-1}]\frac{[\phi^{-1}]^{d}\phi^{d}}{1-\lambda_{d}}[U\sqrt{\chi^D}U^{-1}]g_d
\\&=\frac{P_{d0}}{1-\lambda_d}.\label{Pdallapprox}
\end{split}
\eeq
where in the second line we factor $\bar{P}_{d,\text{tot}}$ into two components, one determined by the BSE's leading eigenvalue $\lambda_d$ and one determined by the pair-mobility $P_{d0}$. The latter is 
given by the non-interacting but dressed two-particle pair-field susceptibility.

$P_{d,\text{tot}}$ and $\bar{P}_{d,\text{tot}}$ are plotted as a function 
of filling in Figs.~\ref{Pddopedepend}(a) and \ref{Pddopedepend}(c), respectively, and the similarity in their density-dependence enables our analysis based on \disp{Pdallapprox}. 
The individual factors $1/(1-\lambda_d)$ and $P_{d0}$ are plotted in \figdisp{Pddopedepend} (b) and \ref{Pddopedepend}(d), respectively. \figdisp{Pddopedepend}(b) shows that $1/(1-\lambda_d)$ diverges more rapidly on the hole doped side as the temperature is lowered. 
The reason why $P_{d,\text{tot}}$ remains larger on the electron-doped side is due to 
the prefactor $P_{d0}$, shown in \figdisp{Pddopedepend}(d); $P_{d0}$ on the electron-doped side is almost twice as large as its value on the hole-doped side. 
This analysis demonstrates that the leading eigenvalue $\lambda_d$ [or $1/(1-\lambda_d$)] is a more suitable proxy for $T_c$, and that conclusions drawn from inspecting the pair-field susceptibility at a temperature too far above $T_c$ can be misleading. 

To test the robustness of our results against finite-size effects, \figdisp{dopedepend4by4} presents a similar analysis of results obtained on a $4\times4$ DCA cluster, where we arrive at a similar set of conclusions. Compared to the results for the $2\times 2$ cluster, here we find that the peaks of the susceptibility in the $4\times4$ cluster [\figdisp{dopedepend4by4}(a)] shift away further from half-filling. This behavior is consistent with the notion that in larger clusters, the maximum $T_c$ shifts to higher doping. Similar to the $2\times 2$ result, \figdisp{dopedepend4by4}(b) again shows that at sufficiently low temperatures, $1/(1-\lambda_d)$ is larger on the hole-doped side than on the electron doped side. These results lend further support to the argument that $\lambda_d$ is a better indicator for the doping dependence of $T_c$.

\begin{figure}[ht]
\includegraphics[width=0.88\linewidth]{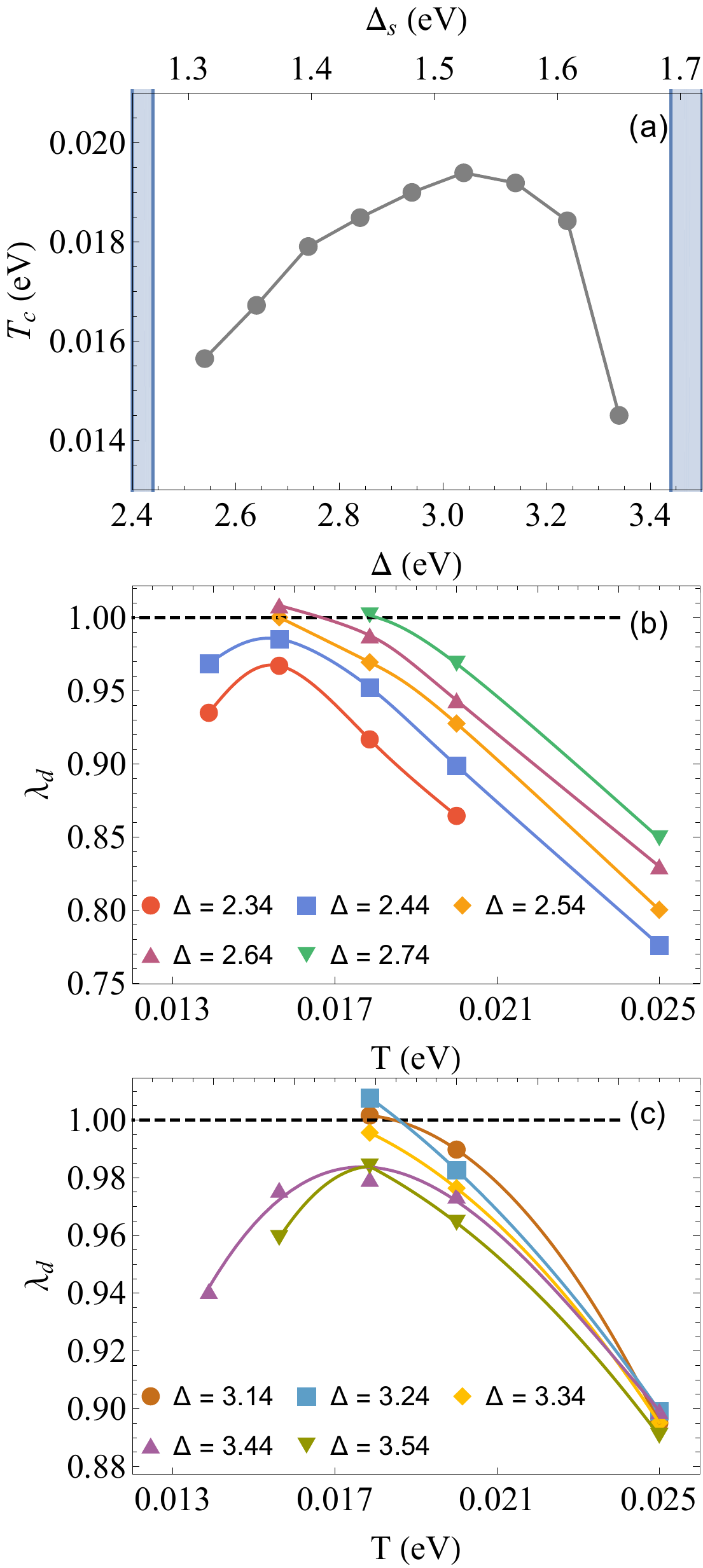}
\caption{(a) Superconducting transition temperature as a function of charge-transfer energy (bottom x-axis) and spectroscopic gap (top x-axis) for a $2\times2$ cluster with $n_h=1.15$. $T_c$ has a peak around $\Delta=3$ eV. For $\Delta \le 2.44$ eV or $\Delta \ge 3.44$ eV, we did not find a finite $T_c$, as marked by the shaded region, and as seen from the temperature dependence of $\lambda_d(T)$ in panels (b) and (c).
}
\label{vsdelta}
\end{figure}

\subsection{Dependence of pair-field susceptibility and transition temperature on the charge-transfer energy} 
Studying the three-band model allows us to account for the material dependence in $T_c$ by considering changes in the charge-transfer energy $\Delta = \varepsilon_p-\varepsilon_d$, which varies between different cuprate superconductors \cite{Ohta,Weber}. Analogous studies in the single-band Hubbard model are challenging because the effective parameters $t$, $t^\prime$, and $U$ all have an implicit dependence on $\Delta$ \cite{Ohta}. 

Figure \ref{vsdelta}(a) plots $T_c$ as a function of the charge-transfer energy, obtained again from a  $2\times2$ cluster at optimal hole-doping $n_h=1.15$. (We reiterate that $T_c$ is determined here by interpolating between the temperatures where $\lambda_d$ crosses one.) We find that $T_c$ has a non-monotonic dependence on $\Delta$, with a maximum occuring near $\Delta \approx 3.1$ eV. Moreover, we do not find any indication of a finite $T_c$ down to the lowest accessible temperatures ($T \sim 0.013$ eV) for $\Delta \le 2.44$ eV or $\Delta \ge 3.44$ eV. For these values of $\Delta$, $\lambda_d$ approaches one upon cooling but eventually turns over indicating competition from another phase [see Figs. \ref{vsdelta}(b) and \ref{vsdelta}(c)].

A recent scanning tunneling microscopy (STM) experiment \cite{Ruan} has correlated the spectroscopic gap $\Delta_s$ in several undoped cuprates and found that it correlates strongly with their optimal $T_c$'s obtained upon doping. To compare with the STM results, we estimated  $\Delta_s$ for a $2\times2$ cluster using exact diagonalization by computing 
the sum of the energy costs of adding a hole to and subtracting a hole from the half-filling system, namely $\Delta_s = E(n_\uparrow=3, n_\downarrow=2)+E(n_\uparrow=1, n_\downarrow=2) - 2E(n_\uparrow=2, n_\downarrow=2)$. 
We found that $\Delta_s$ depends linearly on $\Delta$ in our parameter region with $\Delta_s \approx 0.42\Delta+0.25$. Using this 
relationship, we then mapped our computed $T_c(\Delta)$ onto a $T_c(\Delta_s)$ in \figdisp{vsdelta}(a) (top x-axis). When the transition is present, the experiments show a trend of increasing maximum $T_c$ as $\Delta_s$ decreases~\cite{Ruan}. The $T_c$ in panel (a) for $\Delta_s>1.5$ eV is qualitatively consistent with this trend.

\begin{figure}[ht]
\includegraphics[width=0.92\columnwidth]{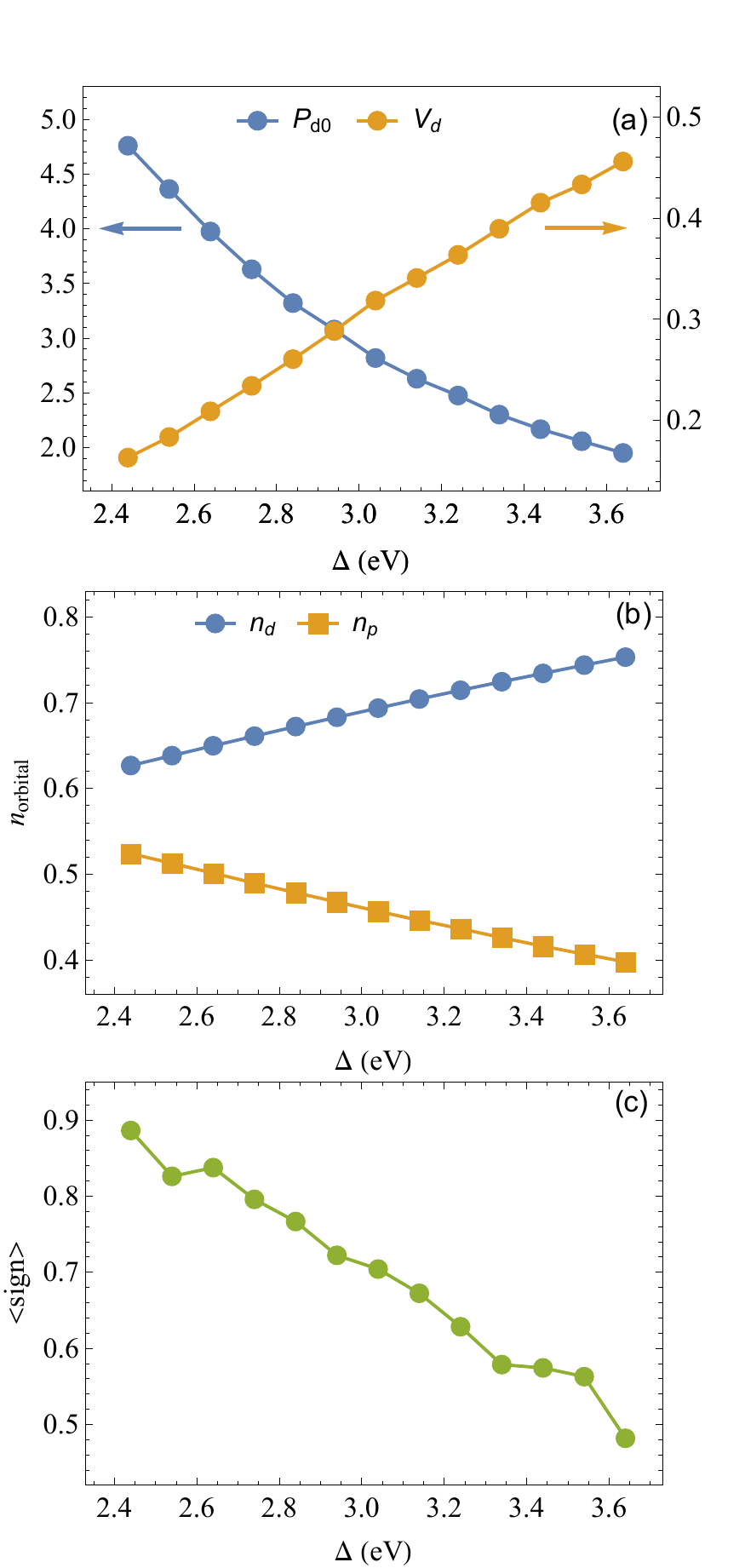}
\caption{(a) The intrinsic pair-field susceptibility $P_{d0}$ and the strength of the $d$-wave pairing interaction, $V_d$, in a separable approximation as a function of the charge transfer gap $\Delta$ for $n=1.15$ and $T=0.025$ eV. The left y-axis is for $P_{d0}$ and the right y-axis is for $V_d$. $P_{d0}$ increases along with $\Delta$ while $V_d$ decreases. (b) and (c) show the $\Delta$ dependence of the orbital density and the average quantum Monte Carlo sign, respectively, at the same temperature for (a).}
\label{Pd0Vd0}
\end{figure}

To analyze this behavior, we then adopted the separable approximation \cite{maier6,maier5}:
\beq\nonumber
\lambda_d(T) \sim P_{d0}(T) V_d(T),
\eeq
 with $P_{d0}$ defined in Eq.~(\ref{Pdallapprox}) and
$V_d(T)$ as the ratio of $\lambda_d(T)$ to $P_{d0}(T)$,
representing the strength of the effective pairing interaction. 
The results for $P_{d0}$ and $V_d$ computed at $T=0.025$ are shown in \figdisp{Pd0Vd0}(a). We find that $P_{d0}$ grows with $\Delta$ while $V_d$ falls monotonically. The competition between these two effects, therefore, gives rise to an optimal $\Delta$ with the highest $T_c$. Given the manageable sign problem in \figdisp{Pd0Vd0}(c) for this region, we believe these results are reliable.

\begin{figure}[ht]
\includegraphics[width=0.92\columnwidth]{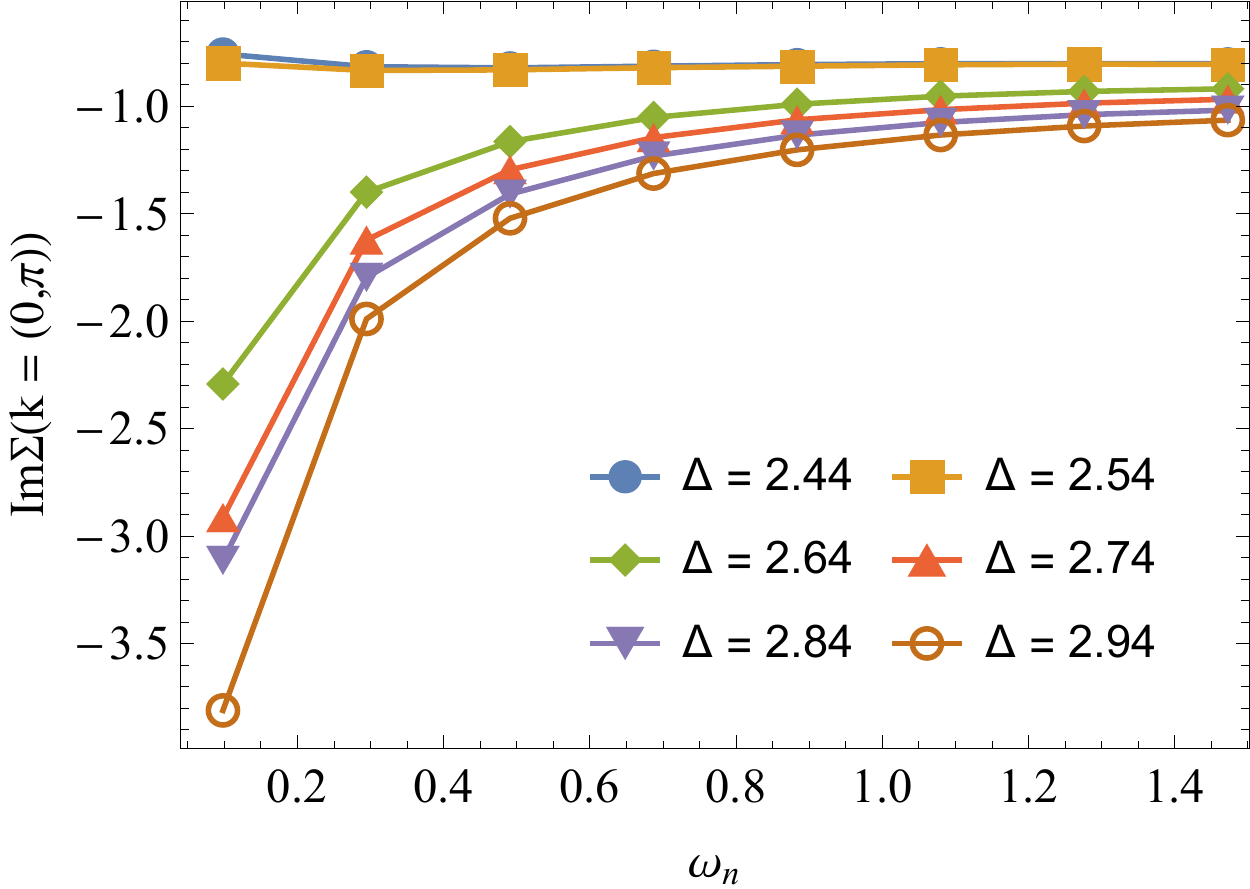}
\caption{$\mathrm{Im}\Sigma({\bf k},\text{i}\omega_n)$ vs $\omega_n$ at ${\bf k}=(0,\pi)$, $T=0.031$, and half-filling for various charge-transfer energy $\Delta$ (in the unit of eV). The metal-insulator transition happens at $\Delta \sim 2.6$.}
\label{imsigma2by2}
\end{figure}

We can gain a more physical understanding of the non-monotonicity of $T_c(\Delta)$ by considering the degree to which the carriers are localized in the three band model. In the three band model, $\Delta$ is the dominant factor setting the charge-transfer energy and increasing this value discourages the doped holes from residing on the oxygen sites, as shown in \figdisp{Pd0Vd0}(b). 
This situation impedes the mobility of electrons since there is no direct Cu-Cu hopping in the model that can bypass the oxygen sites. Conversely, as $\Delta$ reduces, mobility is enhanced, resulting in an increasing pair mobility $P_{d0}$ with decreasing $\Delta$. Further decreasing $\Delta$, the correlation from $U_{dd}$, or the ``Mottness'', is reduced as the carriers can move through the O sublattice \cite{Cooper} and the half-filled system becomes a metal \cite{ZSA} instead of an insulator. We can confirm this behavior in \figdisp{imsigma2by2}, where we plot the imaginary part of the self-energy as a function of Matsubara frequency for ${\bf k}=(\pi,0)$ for the half-filled case for different values of $\Delta$. Here, one sees a transition from insulating behavior at $\Delta \gtrsim 2.6$ eV to metallic behavior at lower $\Delta$, consistent with \refdisp{Vitali}. At the same time, the strength of the effective $d$-wave pairing interaction $V_d$ decreases with decreasing $\Delta$. This behavior can be understood by considering the increased mixing of the uncorrelated O-$p$ states with the correlated Cu-$d$ states at the Fermi level as the charge transfer gap size $\Delta$ decreases. Since $V_d$ is an effective interaction that arises from the Coulomb repulsion $U_{dd}$ on the Cu-$d$ orbitals, this increased orbital distillation leads to a reduction in the effective pairing interaction \cite{Sakakibara}.   

\section{Summary  \& Conclusion}
We have studied the three-band Hubbard model using DQMC and DCA, two complementary non-perturbative methods. Using these approaches, we examined the $d$-wave pairing correlations and superconducting transition temperatures $T_c$ in the model.  
Specifically, we explored the doping dependence of $T_c$ for a $2\times2$ DCA cluster and found a superconducting dome on the hole- ($n_h>1$) and electron-doped ($n_h<1$) sides of the phase diagram. While the hole-doped case presents a higher $T_c$, the pair-field susceptibility $P_{d,\text{tot}}$ above $T_c$ is stronger on the electron-doped side. This result indicates that the  eigenvalue $\lambda_d$ of the particle-particle Bethe-Salpeter equation [or $1/(1-\lambda_d)$] is a better proxy for $T_c$ than the pair-field susceptibility. In particular, we have found that the pair-field susceptibility $P_{d,\text{tot}}$ is too heavily influenced by the bare susceptibility $P_{d0}$ at high temperatures, resulting in a doping dependence qualitatively different from that of $T_c$. Similar behavior was also observed on $4\times4$ clusters, indicating that our results are reasonably robust against finite-size effects.   

Armed with these results, we also examined the effects of the charge-transfer gap size on the superconducting transition temperature $T_c$. We found that there is an optimal charge-transfer gap ($\Delta\approx3.04$ eV in our specific parameter set) that gives rise to a maximum $T_c$. This observation sheds light on how the charge-transfer properties of the cuprates relate to their superconducting transition temperature, thus providing opportunities to further optimize $T_c$.  Moreover, these results are  relevant to the recently discovered superconducting state in the nickelates \cite{Li}, where the charge transfer energy is thought to be much larger \cite{Li, Jiang}. 

\section{Acknowledgements} The authors would like to thank K. Haule, G. Kotliar, H. Terletska, L. Chioncel, A. Georges, S. Karakuzu, P. Dee and E. Huang for useful comments. This work was supported by the Scientific Discovery through Advanced Computing (SciDAC) program funded by the U.S. Department of Energy, Office of Science, Advanced Scientific Computing Research and Basic Energy Sciences, Division of Materials Sciences and Engineering. This research also used resources of the Oak Ridge Leadership Computing Facility, which is a DOE Office of Science User Facility supported under Contract DE-AC05-00OR22725.

\appendix

\section{Components of the $d$-wave pair-field susceptibility}\label{sec:appendix}
\begin{figure}[ht]
\includegraphics[width=0.8\linewidth]{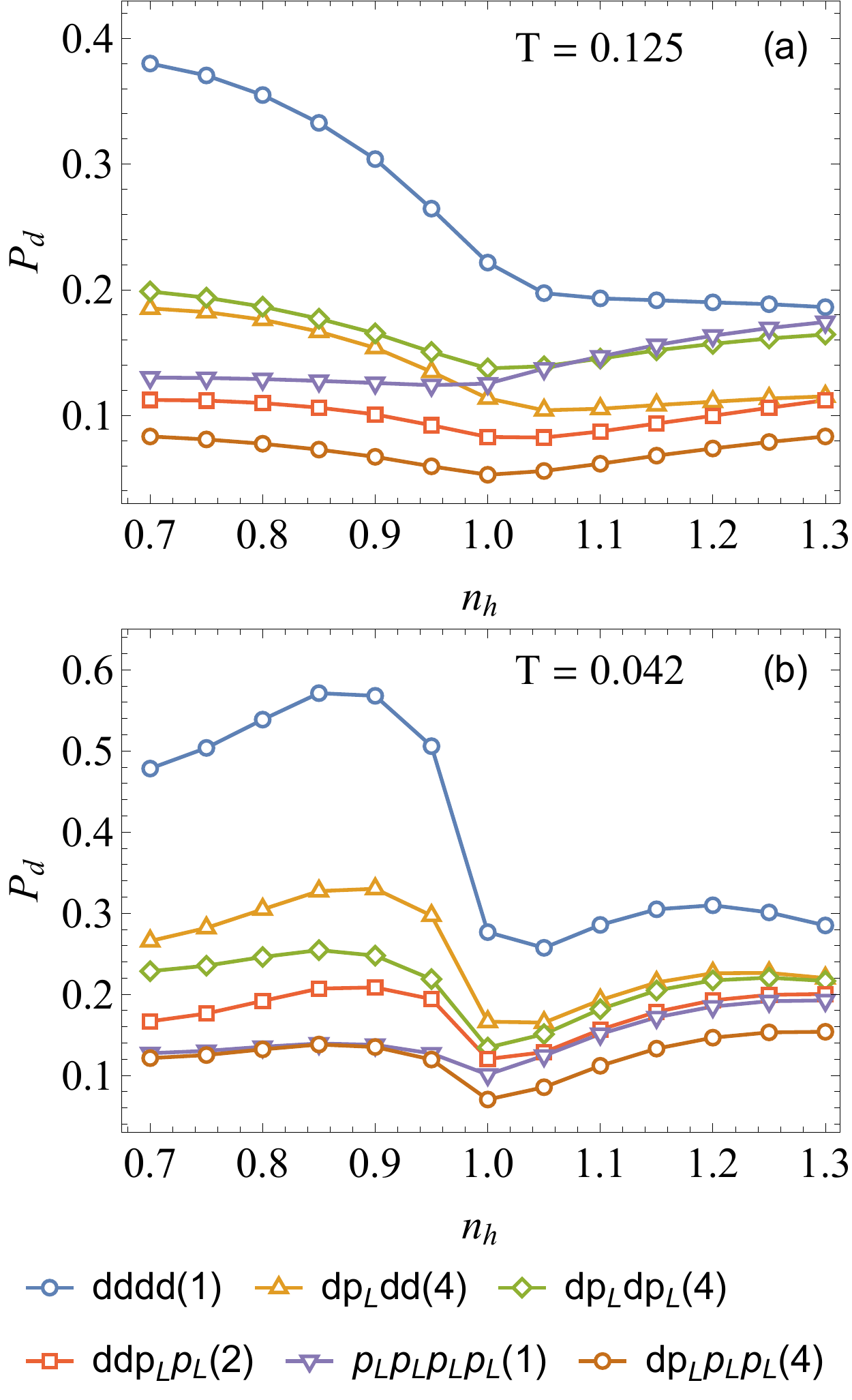}
\caption{ Different components of the $d$-wave pair-field susceptibility as a function of hole density $n_h$ for a $4\times4$ cluster with $U_{pp}=0$ for different temperatures $T=0.125$ (a) and $T=0.042$ eV (b). The number in bracket represents the multiplicity of the same type of component.
}
\label{Pdcomponents}
\end{figure}

For completeness, we plot the different (dominant) components of the total $d$-wave pairfield susceptibility $P_{d,\text{tot}}$ in \figdisp{Pdcomponents}. Each curve exhibits an asymmetry between electron- and hole-doping. The $dddd$ component is the largest individual component, and as temperature decreases, a peak develops on both the hole- and electron-doped sides of the plot. Although the other components are relatively smaller, they could contribute to the total susceptibility significantly considering their multiplicity. For this reason, the $L'$-related components, which are not shown in \figdisp{Pdcomponents} because they are small individually but have large multiplicity, together provide a non-negligible contribution to the total susceptibility especially at higher temperatures.

\end{document}